# Optimal UCA Design for OAM Based Wireless Backhaul Transmission

Haiyue Jing[†], Wenchi Cheng[†], Wei Zhang[‡], and Hailin Zhang[†]
[†]State Key Laboratory of Integrated Services Networks, Xidian University, Xi'an, China
[‡]School of Electrical Engineering and Telecommunications, the University of New South Wales, Sydney, Australia
E-mail: {*hyjing@stu.xidian.edu.cn, wccheng@xidian.edu.cn, w.zhang@unsw.edu.au, hlzhang@xidian.edu.cn*}

*Abstract*—Orbital angular momentum (OAM), which is considered as a novel way to achieve high capacity, has been attracted much attention recently. OAM signals emitted by uniform circular array (UCA) are widely regarded to go through the Bessel-form channels. However, the channel gains corresponding to the Bessel-form channels are with low signal-to-noise-ratio (SNR) on OAM-modes and it is difficult to achieve high capacity using all OAM modes. To achieve maximum capacity offered by OAM multiplexing for wireless backhaul communications, in this paper we propose the optimal UCA design, which selects the optimal OAM-modes and radius of receive UCA. We formulate the capacity maximization problem and divide it into two subproblems for obtaining the corresponding optimal UCA design for OAM multiplexing based wireless backhaul communications. In particular, the optimal radius of the receive UCA is firstly derived. Then, we propose an mode selection scheme to choose appropriate OAM-modes for data transmission to maximize the capacity. Extensive simulations obtained validate that the capacity of OAM multiplexing can be significantly increased with our developed scheme.

*Index Terms*—Orbital angular momentum (OAM) multiplexing, uniform circular array (UCA), Bessel-form channels, mode selection, backhaul.

## I. INTRODUCTION

DURING the past few years, wireless backhaul communications, where the transmitter and receiver are fixed at two locations, have received much attention [1]–[3]. To meet the capacity requirements in wireless backhaul scenarios, the multiplexing techniques such as spatial multiplexing and orthogonal frequency-division multiplexing (OFDM) are used to achieve multiple data streams transmission for increasing the capacity. However, current techniques are difficult to satisfy the increasing capacity demand required by the extremely high-data traffics of wireless backhaul transmission.

Recently, orbital angular momentum (OAM) multiplexing is considered as an another efficient approach to increase the capacity of wireless backhaul communications [4]–[6]. OAM, which is an important property of electromagnetic waves, is a type of wavefront with helical phase [7], [8]. The phase front twists along the propagation direction. The electromagnetic wave carrying OAM contains an infinite number of topological charges, i.e., the OAM-modes. The coaxial waves carrying different OAM-modes are mutually orthogonal with each other. Due to the orthogonality among vorticose beams, multiple data streams can be multiplexed and demultiplexed at the transmitter and receiver without interference, respectively, thus significantly increasing the capacity of wireless backhaul communications.

Because of the flexibility and controllability of digitally generating vorticose beams with multiple OAM-modes simultaneously, uniform circular array (UCA) is considered as a candidate for OAM multiplexing based wireless backhaul communications [9]–[13]. The authors of [9] first demonstrated that UCA can generate OAM-modes by simulation. The authors of [11] experimentally concluded that OAM multiplexing based UCA can be used within a certain distance range. Since the high-order OAM-modes are difficult to be used for data transmission in the $X$-frequency, the authors of [10] designed the UCA configuration to generate high-order OAM-modes with high quality, thus providing guarantee for OAM multiplexing. The authors of [3], [12] mathematically analyzed the OAM multiplexing based UCA in backhaul scenarios, showing that OAM multiplexing can achieve high capacity and there is a vast reduction in receiver complexity of demultiplexing OAM-modes. To further increase the capacity, the authors of [13] proposed OAM-embedded-multiple-input-multiple-output (MIMO) communication framework to obtain the capacity of joint OAM multiplexing and MIMO.

In the research on OAM multiplexing based UCA, the channel model is usually considered to follow the Bessel forms with respect to wavelength, OAM-modes, transmission distance, and the radii of transmit and receive UCAs [11], [13]–[15]. The authors of [11] only analyzed the impact of transmission distance on OAM multiplexing. The authors of [14] discussed the channel gain in a special case, where the channel gain determined by the order of OAM-mode decreases as the order of OAM-mode increases. However, in terms of the characteristics corresponding to Bessel function, the channel gains corresponding to different OAM-modes are oscillated and the radii of transmit and receive UCAs have great influence on the channel gain. There are few published papers taking the impact of the radii corresponding to transmit and receive UCAs on capacity into consideration in backhaul scenarios and it remains an open question.

Motivated by the above-mentioned problem, in this paper we focus on the impact of Bessel-forms channels to achieve the maximum capacity with fixed transmission distance in backhaul wireless communications. Since the impacts of the

This work was supported in part by the National Natural Science Foundation of China (No. 61771368) and the Young Elite Scientists Sponsorship Program by CAST (2016QNRC001).

radii corresponding to transmit and receive UCAs are the same, the impact of the radius of receive UCA is only analyzed. We propose the optimal UCA design to select the optimal OAM-modes and radius of receive UCA for increasing the capacity of OAM multiplexing based UCA in wireless backhaul communications. To obtain the optimal UCA design for achieving the maximum capacity, we divide the capacity maximization problem into two subproblems. We propose the mode selection scheme to select the OAM-modes for data transmission by a given threshold. The obtained numerical results show that the OAM multiplexing with our proposed optimal UCA design can significantly increase the capacity as compared with that without the proposed optimal UCA design.

The rest of this paper is organized as follows. Section II gives system model for OAM multiplexing based UCA in wireless backhaul communications. Section III derives the channel model for OAM multiplexing based UCA, proposes the optimal UCA design to increase the capacity. Section IV derives the optimal threshold and the radius of receive UCA to maximize the capacity. Section V evaluates our proposed schemes and compares the capacity of our proposed schemes with that without our proposed schemes. The paper concludes with Section VI.

## II. SYSTEM MODEL FOR OAM MULTIPLEXING BASED WIRELESS BACKHAUL COMMUNICATIONS

Figure 1 depicts the system model for OAM multiplexing based UCA in wireless backhaul communications, where both transmitter and receiver are UCAs and they are aligned with each other. There are $N$ and $M$ antenna elements equipped on the transmit and receive UCA, respectively. The gaps of position angles between two adjacent antenna elements on the transmit and receive UCAs are $2\pi/N$ and $2\pi/M$, respectively. For the transmit UCA, the antenna elements, uniformly around the perimeter of the circle, are fed with the same input signal but with a successive delay from antenna element to antenna element such that after a full turn the phase has been incremented by $2\pi l$, where $l$ represents the number of topological charges, i.e., OAM-modes. For the receive UCA, the antenna elements are also uniformly around perimeter of the circle. The notations $R_t$ and $R_r$ denotes the radii of transmit and receive UCAs, respectively. We denote by $d$ the distance between the center of the transmit UCA and the center of the receive UCA. The parameter $d_{mn}$ is defined as the distance from the $n$th ($1 \leq n \leq N$) antenna element on the transmit UCA to the $m$th ($1 \leq m \leq M$) antenna element on the receive UCA. The parameter $\alpha$ denotes the angle between $x$-axis and the first antenna element on the receive UCA.

For OAM multiplexing based UCA in wireless backhaul communications, the OAM-modes are selected to use for data transmission based on the threshold $\varrho$, which depends on the channel state information. The input signals becomes twisted using the inverse discrete Fourier transform (IDFT) and then, the vorticose signals carrying the selected OAM-modes are emitted by the transmit UCA. At the receiver, the radius of receive UCA can change to efficiently receive the signals. The receive signals are decomposed using the discrete Fourier transform (DFT). In the following, we propose the optimal UCA design and the mode selection scheme for for wireless backhaul communications. Also, the achieved capacity using our proposed schemes is analyzed. Note that our proposed schemes apply to not only the scenario where the transmit and receive UCA are aligned with each other perfectly, but also the scenario where the transmit and receive UCAs are misaligned with each other.

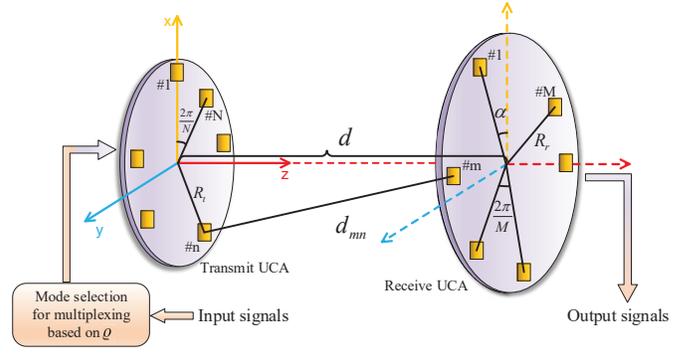

Fig. 1. The system model for the OAM multiplexing based wireless backhaul communications.

## III. CHANNEL MODEL AND PROPOSED SCHEMES FOR WIRELESS BACKHAUL COMMUNICATIONS

The transmit UCA equipped with $N$ antenna elements can generate $N$ OAM-modes simultaneously [16]. The transmit signal of the $n$th antenna element on the transmit UCA, denoted by $x_n$, is given as follows:

$$x_n = \sum_{l=0}^{N-1} \frac{1}{\sqrt{N}} p_l s_l e^{j\varphi_n l} = \sum_{l=0}^{N-1} \frac{1}{\sqrt{N}} p_l s_l e^{j\frac{2\pi(n-1)}{N}l}, \quad (1)$$

where $\varphi_n = \frac{2\pi(n-1)}{N}$ denotes the azimuthal angle, defined as the angular position with respect to the $n$th antenna element on the transmit UCA, $l$ ($0 \leq l \leq N-1$) represents the order of OAM-mode, $1/\sqrt{N}$ is the normalization factor, $s_l$ represents the input symbol corresponding to the $l$th OAM-mode, and $p_l$ is denoted by the power allocation factor corresponding to the $l$th OAM-mode. We denote by $\boldsymbol{x} = [x_1, x_2, \cdots, x_N]^T$ the transmit signal vector, where $(\cdot)^T$ represents the transpose operation. The transmit signal vector $\boldsymbol{x}$ can be written as follows:

$$\boldsymbol{x} = \boldsymbol{W}\boldsymbol{P}\boldsymbol{s}, \quad (2)$$

where the notation $\boldsymbol{W}$ is the IDFT matrix with elements $\frac{1}{\sqrt{N}}\exp\left[j\frac{2\pi}{N}(n-1)l\right]$ and $\boldsymbol{W}$ is an unitary matrix. The parameter $\boldsymbol{P} = \text{diag}(p_0, p_1, \cdots, p_{N-1})$ is a diagonal matrix. The vector $\boldsymbol{s} = [s_0, s_1, \cdots, s_{N-1}]^T$ and $\mathbb{E}\{\boldsymbol{s}^H\boldsymbol{s}\} = 1$, where $\mathbb{E}\{\cdot\}$ represents the expectation operation and $(\cdot)^H$ denotes the conjugate transpose operation.

For wireless backhaul communications, the channel gain, denoted by $h_{mn}$, from the $n$th antenna element at the transmitter to the $m$th antenna element at the receiver is given as follows: [4]

$$h_{mn} = \frac{\beta\lambda e^{-j\frac{2\pi}{\lambda}d_{mn}}}{4\pi d_{mn}}, \quad (3)$$

where $\lambda$ represents the wavelength. The notation $\beta$ denotes all relevant constants such as attenuation and phase rotation caused by antennas and their patterns on both sides. The distance $d_{mn}$ is derived as follows:

$$d_{mn} = \sqrt{d^2 + R_t^2 + R_r^2 - 2R_tR_r\cos(\varphi_n - \phi_m - \alpha)}, \quad (4)$$

where $\phi_m = 2\pi(m-1)/M$ is the basic rotation-angle for the $m$th receive antenna element and $\phi_m + \alpha$ is the azimuthal angle, which is defined as the angular position corresponding to the $m$th antenna element on the receive UCA.

Note that only the number of antenna elements on the receive UCA is equal to or larger than the number of antenna elements on the transmit UCA, i.e., $M \geq N$, can the input symbols be decomposed effectively. To make full use of antenna resources, in this paper we equate the number of antenna elements at the receiver to that at the transmitter, i.e., $M = N$. Thus, the channel matrix, denoted by $\boldsymbol{H} = (h_{mn})_{N \times N}$, is a circulant matrix due to the item $\cos(\varphi_n - \phi_m - \alpha)$ ($1 \leq n, m \leq N$).

Since the channel matrix $\boldsymbol{H}$ is a circualnt matrix, we have

$$\boldsymbol{H} = \boldsymbol{W}\boldsymbol{\Upsilon}\boldsymbol{W}^H, \quad (5)$$

where $\boldsymbol{\Upsilon}$ is a diagonal matrix. The diagonal elements of $\boldsymbol{\Upsilon}$ are the eigenvalues, which are the $N$-point DFT of the first row corresponding to $\boldsymbol{H}$. Since the singular values of $\boldsymbol{H}$ can be derived by calculating the modulus with respect to the diagonal elements of $\boldsymbol{\Upsilon}$, the $l$th singular value related to $\boldsymbol{H}$, denoted by $\gamma_l$ ($0 \leq l \leq N-1$), is derived as follows:

$$\gamma_l = \left| \sum_{n=1}^{N} h_{1n} \exp\left[-j\frac{2\pi}{N}(n-1)l\right] \right|. \quad (6)$$

In Eq. (6), $\gamma_l$ can be considered as the channel amplitude gain corresponding to the $l$th OAM-mode.

Then, the receive signal vector, denoted by $\boldsymbol{y} = [y_1, y_2, \cdots, y_N]^T$, is derived as follows:

$$\boldsymbol{y} = \boldsymbol{H}\boldsymbol{x} + \boldsymbol{z} = \boldsymbol{H}\boldsymbol{W}\boldsymbol{P}\boldsymbol{s} + \boldsymbol{z}, \quad (7)$$

where $\boldsymbol{z} = [z_1, z_2, \cdots, z_N]^T$ is the noise vector and $z_m$ is the noise received by the $m$th antenna element.

To achieve the single-symbol detection, we multiply the DFT matrix $\boldsymbol{W}^H$ with the receive signal vector as follows:

$$\boldsymbol{W}^H\boldsymbol{y} = \boldsymbol{W}^H\boldsymbol{H}\boldsymbol{x} + \boldsymbol{W}^H\boldsymbol{z} = \boldsymbol{\Upsilon}\boldsymbol{P}\boldsymbol{s} + \boldsymbol{W}^H\boldsymbol{z}. \quad (8)$$

Since $\boldsymbol{W}^H$ is an unitary matrix, the noise statistics does not change. Following Eq. (8), the input symbols can be easily recovered. Then, the received channel signal-to-noise-ratio (SNR), denoted by $SNR$, can be derived as follows:

$$SNR = \frac{\overline{P}\boldsymbol{P}^2}{|\boldsymbol{W}^H\boldsymbol{z}|^2 |\boldsymbol{\Upsilon}^{-1}|^2} = \frac{\overline{P}|\boldsymbol{\Upsilon}|^2\boldsymbol{P}^2}{\sigma^2}, \quad (9)$$

where $\sigma^2$ represents the variance of receive noise and $\overline{P}$ is the total power.

For the OAM multiplexing based UCA system, the capacity, denoted by $C_s$, can be derived as follows:

$$C_s = B\log_2(1 + SNR) = B\log_2\left(1 + \frac{|\boldsymbol{\Upsilon}|^2\boldsymbol{P}^2\overline{P}}{\sigma^2}\right)$$
$$= \sum_{l=0}^{N-1} B\log_2\left(1 + \frac{\gamma_l^2 p_l^2 \overline{P}}{\sigma^2}\right), \quad (10)$$

where the notation $B$ represents the bandwidth. Given the bandwidth $B$, the noise variance $\sigma^2$, and the power allocation factor $p_l$ corresponding to the $l$th OAM-mode, the capacity $C_s$ is determined by the subchannel amplitude gain $\gamma_l$ regarding each OAM-mode and the number of subchannels used for data transmission. If all OAM-modes are used to convey the information, there exists power waste because subchannel amplitude gains corresponding to some OAM-modes are very small, resulting in very low SNR.

Referring to Eq. (15) in [16], when the number of antenna elements is relatively large, we can derive subchannel amplitude gain as follows:

$$\gamma_l = \left| \frac{\beta\lambda\sqrt{N}e^{-j\frac{2\pi}{\lambda}\sqrt{d^2+R_t^2+R_r^2}}e^{j\alpha l}}{4\pi dj^l} J_l\left(\frac{2\pi R_t R_r}{\lambda\sqrt{d^2+R_t^2+R_r^2}}\right) \right|. \quad (11)$$

In view of the item $J_l\left(\frac{2\pi R_t R_r}{\lambda\sqrt{d^2+R_t^2+R_r^2}}\right)$, the capacity $C_s$ depends on the radii of transmit and receive UCAs with fixed wavelength $\lambda$, the distance $d$, and the order of OAM-mode $l$. Since the impact of $R_t$ on capacity $C_s$ is the same as that of $R_r$ on capacity $C_s$, we only discuss the impact of $R_r$. The value of bessel function related to the $l$th OAM-mode depends on the item $\frac{2\pi R_t R_r}{\lambda\sqrt{d^2+R_t^2+R_r^2}}$ and the value of bessel function is oscillated, thus resulting in low capacity and waste of power in some cases. To reduce the impact of bessel function and make full use of power for increasing capacity, we propose the optimal UCA design to select OAM-modes used for data transmission and adjust the radius of receive UCA for efficiently receiving the vorticose signals carrying the selected OAM-modes.

To select appropriate OAM-modes corresponding to high channel gains for conveying information, we propose an mode selection scheme. The threshold $\varrho$ is set to decide which OAM-mode can be used for data transmission. For example, when the value of the item $J_l\left(\frac{2\pi R_t R_r}{\lambda\sqrt{d^2+R_t^2+R_r^2}}\right) \geq \varrho$, the corresponding OAM-mode can be selected for data transmission. When the value of the item $J_l\left(\frac{2\pi R_t R_r}{\lambda\sqrt{d^2+R_t^2+R_r^2}}\right) < \varrho$, the corresponding OAM-mode cannot be selected for data transmission. After OAM-mode selection, the transmit UCA can emit the data streams corresponding to the selected OAM-modes instead of all OAM-modes.

## IV. CAPACITY OPTIMIZATION FOR OAM MULTIPLEXING BASED UCA IN WIRELESS BACKHAUL COMMUNICATIONS

### A. Mode Selection Scheme

For the mode selection scheme, the value of $\varrho$ has impact on the capacity and different $\varrho$ correspond to different capacities.

Specially, the power is averagely allocated to the selected OAM-mode. To derive $\varrho$ for capacity maximization, we set $a(l)$ as follows:

$$a(l) = \frac{1}{2} + \frac{1}{2}\text{sgn}\left[J_l\left(\frac{2\pi R_t R_r}{\lambda\sqrt{d^2 + R_t^2 + R_r^2}}\right) - \varrho\right], \quad (12)$$

where $a(l) = 1$ represents the $l$th OAM-mode is selected to convey information and $a(l) = 0$ represents the $l$th OAM-mode cannot be used for data transmission. Then, the capacity $C_s$ with respect to $\varrho$ can be derived as follows:

$$C_s(\varrho) = \sum_{l=0}^{N-1} B\log_2\left(1 + \frac{a(l)\gamma_l^2\overline{P}}{\sigma^2 \sum_{l=0}^{N-1} a(l)}\right), \quad (13)$$

where $\sum_{l=0}^{N-1} a(l)$ represents the number of selected OAM-modes which can be used for data transmission. To achieve the maximum capacity using our proposed mode selection scheme, the developed algorithm to find the optimal threshold $\varrho$ is presented in **Algorithm 1**.

### B. Optimal UCA Design

With fixed $R_t$ and $\lambda$, the capacity of OAM multiplexing based UCA system can be maximized by selecting optimal threshold $\varrho$ and controlling the radius of receive UCA. Consequently, the capacity maximization problem, denoted by **P1**, can be formulated as follows:

$$\textbf{\textit{P1:}} \max\ C_s(\varrho, R_r) = \sum_{l=0}^{N-1} B\log_2\left(1 + \frac{a(l)\gamma_l^2\overline{P}}{\sigma^2 \sum_{l=0}^{N-1} a(l)}\right) \quad (14)$$

$$\text{s.t.}: 1). \ R_{\min} \leq R_r \leq R_{\max}; \quad (15)$$
$$2). \ \varrho > 0. \quad (16)$$

In **P1**, $R_{\min}$ and $R_{\max}$ are denoted by the minimum and maximum radius of the receive UCA, respectively. It is clear that **P1** is a constrained optimization problem and optimal solutions exist. Since the threshold $\varrho$ is decided by the values of $J_l\left(\frac{2\pi R_t R_r}{\lambda\sqrt{d^2 + R_t^2 + R_r^2}}\right)$ corresponding to OAM-modes and $J_l\left(\frac{2\pi R_t R_r}{\lambda\sqrt{d^2 + R_t^2 + R_r^2}}\right)$ depends on the radius of receive UCA $R_r$, the optimal solution related to **P1** can be divided into two subproblems to derive the optimal radius of receive UCA and the optimal threshold $\varrho$, respectively, in the following.

**Subproblem 1:** The capacity maximization problem with respect to $R_r$, denoted by **P2**, can be first solved. Problem **P2** is given as follows:

$$\textbf{\textit{P2:}} \max\ C_s(R_r) = \sum_{l=0}^{N-1} B\log_2\left(1 + \frac{\gamma_l^2 p_l^2 \overline{P}}{\sigma^2}\right) \quad (17)$$

**Algorithm 1** : An Interpolation Method for Finding $\varrho$

1: $u = \min\left(\left\{\left|J_l\left(\frac{2\pi R_t R_r}{\lambda\sqrt{d^2+R_t^2+R_r^2}}\right)\right|, 0 \leq l \leq N-1\right\}\right)$
   the minimum initial boundary
   $v = \max\left(\left\{\left|J_l\left(\frac{2\pi R_t R_r}{\lambda\sqrt{d^2+R_t^2+R_r^2}}\right)\right|, 0 \leq l \leq N-1\right\}\right)$
   the maximum initial boundary
2: $C = 0$
3: $w_2 = \frac{u+v}{2}; w_1 = \frac{u+w_2}{2}; w_3 = \frac{w_2+v}{2}$
4: Calculate the capacities $C_s(w_1)$, $C_s(w_2)$, and $C_s(w_3)$.
5: **if** $C_s(w_1) = \max\{C_s(w_1), C_s(w_2), C_s(w_3)\}$ **then**
6:   $v = w_2$;
7:   **if** $C \neq C_s(w_1)$ **then**
8:     $C = C_s(w_1)$; Go to **Step 3**;
9:   **else**
10:    Return $\varrho = w_1$;
11:  **end if**
12: **else if** $C_s(w_2) = \max\{C_s(w_1), C_s(w_2), C_s(w_3)\}$ **then**
13:   $u = w_1$ and $v = w_3$;
14:   **if** $C \neq C_s(w_2)$ **then**
15:     $C = C_s(w_2)$; Go to **Step 3**;
16:   **else**
17:     Return $\varrho = w_2$;
18:   **end if**
19: **else if** $C_s(w_3) = \max\{C_s(w_1), C_s(w_2), C_s(w_3)\}$ **then**
20:   $u = w_2$;
21:   **if** $C \neq C_s(w_3)$ **then**
22:     $C = C_s(w_3)$; Go to **Step 3**;
23:   **else**
24:     Return $\varrho = w_3$;
25:   **end if**
26: **end if**

$$\text{s.t.}: \ R_{\min} \leq R_r \leq R_{\max}. \quad (18)$$

The Lagrangian function for **P2**, denoted by $\mathcal{J}$, is given as follows:

$$\mathcal{J}(R_r, \mu, \nu) = C_s(R_r) + \mu(R_r - R_{\min}) + \nu(R_{\max} - R_r), \quad (19)$$

where $\mu$ and $\nu$ are the nonnegative Lagrangian multipliers corresponding to Eq. (18). Taking the derivative for $\mathcal{J}$ with respect to $R_r$, we can obtain the necessary conditions for optimal radius of the receive UCA, i.e., $karush-kuhn-Tucker$ (KKT) conditions as follows:

$$\begin{cases} \frac{\partial \mathcal{J}(R_r, \mu, \nu)}{\partial R_r} = \frac{dC_s(R_r)}{dR_r} + \mu - \nu = 0; & (20a) \\ \mu(R_r - R_{\min}) = 0; & (20b) \\ \nu(R_{\max} - R_r) = 0. & (20c) \end{cases}$$

where $\frac{dC_s(R_r)}{dR_r}$ is given by Eq. (21).

$$\frac{dC_s(R_r)}{dR_r} = \sum_{l=0}^{N-1} \frac{\frac{2\pi R_t(d^2+R_t^2)}{\lambda(d^2+R_t^2+R_r^2)^{\frac{3}{2}}} J_l\left(\frac{2\pi R_t R_r}{\lambda\sqrt{d^2+R_t^2+R_r^2}}\right)\left[J_{l-1}\left(\frac{2\pi R_t R_r}{\lambda\sqrt{d^2+R_t^2+R_r^2}}\right) - J_{l+1}\left(\frac{2\pi R_t R_r}{\lambda\sqrt{d^2+R_t^2+R_r^2}}\right)\right]}{\frac{16\pi^2\sigma^2}{p_l^2\overline{P}\beta^2\lambda^2 N^2} + J_l^2\left(\frac{2\pi R_t R_r}{\lambda\sqrt{d^2+R_t^2+R_r^2}}\right)}. \quad (21)$$

To obtain Eq. (21), the characteristic of Bessel function is used and it is given as follows:

$$2\frac{dJ_l(x)}{dx} = J_{l-1}(x) - J_{l+1}(x). \quad (22)$$

Based on the KKT conditions and the range of Lagrangian multipliers $\mu$ and $\nu$, three cases are taken into consideration to derive the optimal radius of the receive UCA $R_r$ corresponding to Eq. (20).

**Case I** ($\mu = 0, \nu \neq 0$): If $\mu = 0$ and $\nu \neq 0$ hold, $R_r = R_{\max}$. Substituting $R_{\max}$ into Eq. (21), $R_{\max}$ is the optimal radius of receive UCA if $\frac{dC_s(R_r)}{dR_r}\big|_{R_r=R_{\max}} > 0$.

**Case II** ($\mu \neq 0, \nu = 0$): If $\mu \neq 0$ and $\nu = 0$ hold, $R_r = R_{\min}$. Substituting $R_{\min}$ into Eq. (21), we can obtain the optimal radius of receive UCA is $R_{\min}$ if $\frac{dC_s(R_r)}{dR_r}\big|_{R_r=R_{\min}} < 0$.

**Case III** ($\mu = 0, \nu = 0$): If $\mu = 0$ and $\nu = 0$ hold, the optimal radius of receive UCA is the root of $\frac{dC_s(R_r)}{dR_r} = 0$.

**Subproblem 2:** After deriving the optimal radius of receive UCA, the capacity can achieve maximization using our proposed mode selection scheme. Using **Algorithm 1**, we can find the threshold $\varrho$ and then, the maximum capacity corresponding to **P1** is achieved.

## V. PERFORMANCE OF OAM MULTIPLEXING BASED WIRELESS BACKHAUL COMMUNICATIONS

In this section, we numerically evaluate OAM multiplexing based UCA in wireless backhaul communications with our proposed optimal UCA design. First, we give the optimal radius of the receive UCA corresponding to different numbers of antenna elements and transmission distances. Then, we compare the capacities of OAM multiplexing using our proposed optimal UCA design with that of OAM multiplexing without using the optimal UCA design. We also evaluate the utilization rates of our proposed schemes. Throughout our evaluations, we set the wavelength $\lambda$, the constant $\beta$, and the bandwidth $B$ as 0.1 m, 1, and 20 MHz, respectively.

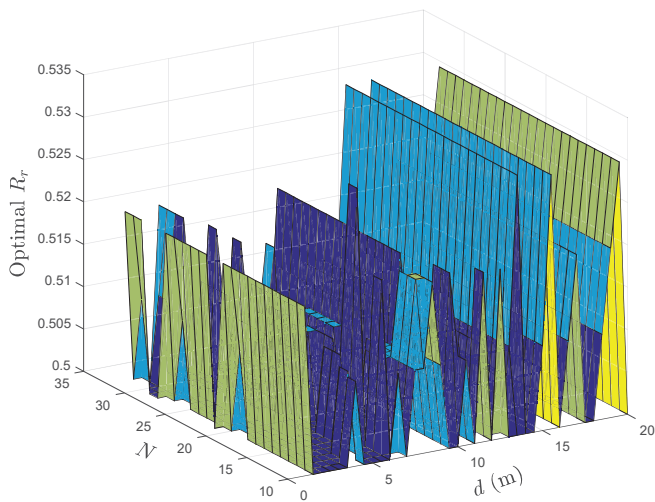

Fig. 2. Optimal $R_r$ for OAM multiplexing with the optimal UCA design in wireless backhaul communications.

Figure 2 shows the optimal radius of the receive UCA $R_r$ for OAM multiplexing in wireless backhaul communications versus the number of antenna elements $N$ and the transmission distance $d$, where we set the radius of transmit UCA $R_t = 0.5\lambda$. We can observe that when the transmission distance $d$ is small, the number of antenna elements has a little impact on the optimal $R_r$ and different numbers of antenna elements correspond to different optimal $R_r$. When the transmission distance $d$ is larger than 14 m, there is not impact of the number of antenna elements on the optimal $R_r$. This is because that the channel gains corresponding to the selected OAM-modes used for conveying data have little changes as the number of antenna elements increases. With fixed number of antenna elements, the optimal $R_r$ randomly varies instead of linearly increasing as the transmission distance $d$ increases. This is because that the Bessel functions corresponding to different OAM-modes are oscillated.

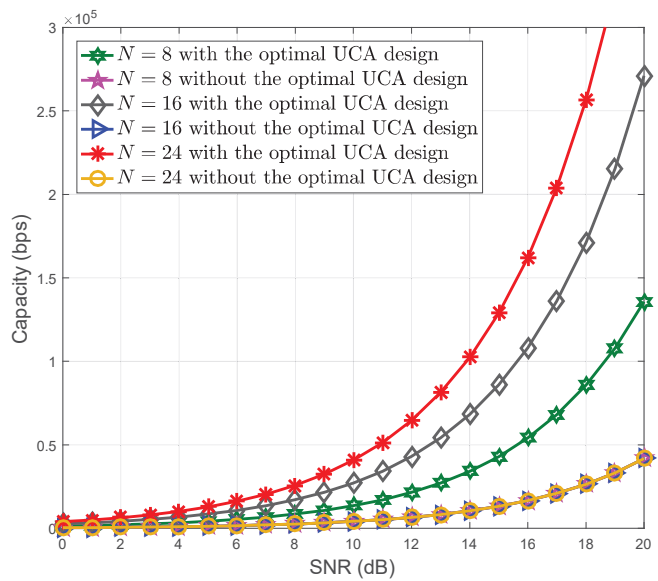

Fig. 3. Capacities for OAM multiplexing with our proposed optimal UCA design and OAM multiplexing without our proposed optimal UCA design in wireless backhaul communications versus SNR.

Figure 3 displays the capacities for OAM multiplexing with our proposed optimal UCA design and OAM multiplexing without our proposed optimal UCA design in wireless backhaul communications versus SNR, where the radius of transmit UCA $R_t = 10\lambda$ and the transmission distance $d = 20$ m. We can obtain that the capacity of OAM multiplexing with the optimal UCA design is larger than that of OAM multiplexing without our proposed optimal UCA design. This is because the radius of receive UCA $R_r$ corresponding to the maximum capacity is applied and the power allocated to the selected OAM-modes with the mode selection scheme increases as compared with the power allocated to all the OAM-modes. Also, the capacities of OAM multiplexing with our proposed optimal UCA design increase as the number of OAM-modes increases. This is because the number of OAM-modes selected by the mode selection scheme increases as the number of OAM-modes. The capacity of OAM multiplexing without our proposed optimal UCA design keeps unchange as the number of OAM-modes increases. This is because the number of OAM-modes which corresponds to high channel amplitude gains doesn't increase and the SNRs of each OAM-modes corresponding to different number of antenna elements are the same.

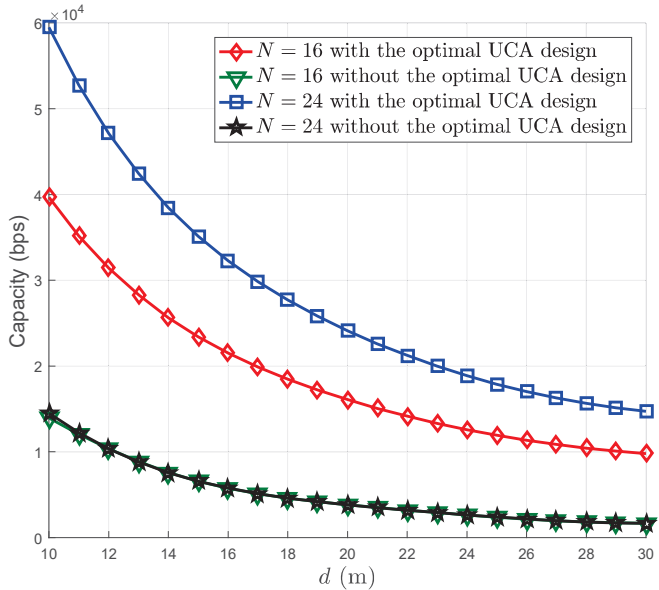

Fig. 4. Capacities for OAM multiplexing with our proposed optimal UCA design and OAM multiplexing without our proposed optimal UCA design in wireless backhaul communications versus transmission distance $d$.

Figure 4 depicts the capacities for OAM multiplexing with our proposed optimal UCA design and OAM multiplexing without our proposed optimal UCA design in wireless backhaul communications versus transmission distance $d$, where the radius of transmit UCA $R_t = 5\lambda$ and the power $P = 10$ dB. The capacity of OAM multiplexing in wireless backhaul communications decreases as the transmission distance $d$ increases. Results coincide with the fact that the receive SNR decreases as the transmission distance increases. The capacity of OAM multiplexing in wireless backhaul communications increases as the number of antenna elements increases. This is because the number of selected OAM-modes increases as the number of antenna elements increases. The capacity of OAM multiplexing with our proposed optimal UCA design is much larger than that of OAM multiplexing without our proposed optimal UCA design. This is because more power allocated to the selected OAM-modes using the optimal threshold and the optimal radius of receive UCA is applied. The gap between the capacities regarding different number of OAM-modes for OAM multiplexing decreases as the transmission distance $d$ increases. This is because the number of selected OAM-modes used for data transmission decreases and the channel gains decrease severely as the transmission distance $d$ increases.

## VI. Conclusions

In this paper, we proposed the optimal UCA design to determine the optimal threshold at the transmitter and the optimal radius of the receive UCA at the receiver for OAM multiplexing based UCA in wireless backhaul communications. To evaluate the performance, we transformed the capacity maximization problem into two equivalent subproblems to derive the optimal radius of receive UCA and the optimal threshold using our proposed mode selection scheme. Simulation results validated that different transmission distances should match with different radii of receive UCAs to achieve the maximum capacity. OAM multiplexing with our proposed optimal UCA design can significantly increase the capacity as compared with OAM multiplexing without our proposed optimal UCA design in wireless backhaul communications.